# Resistive-Switching Dynamics in Poly(3-hexylthiophene-2,5-diyl) Thin Films under Perforated Bottom Electrode

*Sirsendu Ghosh and Pramod Kumar\**

Department of Physics, Indian Institute of Technology Bombay, Mumbai, Maharashtra, 400076, India

*E-mail: pramod_k@iitb.ac.in

**Abstract:**

In this paper, we demonstrate the resistive switching (RS) mechanism in organic semiconductor (OSC), Poly(3-hexylthiophene-2,5-diyl) (P3HT), due to the presence of the perforated bottom electrode (PBE). Multiple types of RS have been observed in P3HT due to the presence of the PBE. The simulation shows a high local electric field at the edges of a patterned bottom electrode (BE), which can increase the probability of metal filament formation due to high current density, suggesting that the use of a PBE can assist the RS mechanism. RS involves switching from the high resistive state (HRS) to the low resistive state (LRS) known as the 'SET' process at higher positive bias, and returning to HRS from LRS is known as the 'RESET' process, which can be achieved at a negative bias. Various switching mechanisms are segregated from each other by the obtained current response to applied voltage. RS due to the formation of complete metal filaments between the top and bottom electrodes showed Ohm's law behaviour. On the other hand, a slope of ~2 in the log-log plot signifies that the space charge limited current (SCLC) dominates the device, and hence RS comes from incomplete metal filament formation or some changes in the P3HT polymer itself. The incomplete metal filament formation can reduce the conduction path length, which results in the RS mechanism. Similarly, high current density can transform of molecular arrangement from crystalline to amorphous state due to joule heating, which leads to an intermediate OFF state. The high current density joule heating RS are from HRS to LRS, which is opposite to the metal filament based RS, hence it is called an inverted RS. The formation of an amorphous state in the polymer is observed by a blue shift in the absorption spectrum achieved at higher temperature that signifies the transformation to an amorphous state. This intermediate OFF/HRS state returns to the ON/LRS state while backwards sweeping from the high positive voltage, ensuring that the charges again follow the crystalline domains while maintaining the SCLC mechanism. The optical images of the fresh device and after multiple cycles indicate the metal percolation inside the OSC responsible for the RS. EDX spectrum at LRS in a cross-sectional transmission electron microscope (TEM) confirms the top metal percolation through the OSC and touches the BE. Therefore, the metal filament formation is the fundamental reason for these observed switching behaviours in P3HT. Physics behind the different types of RS on the same OSC might give a new insight into the memory and neuromorphic computing area.

## 1. Introduction:

Resistive switching (RS) based on perovskites [1][2], binary oxides [3], solid electrolytes [4][5], chalcogenides [6], and organic semiconductors (OSCs) [7][8] has been well reported. There are already many established theories to explain the RS, such as filament formation, conformational change [9], charge trapping/detrapping [10][11], electrical barrier modulation [12] etc. The OSCs come with their own advantages of easy fabrication process through solution processing, molecular design flexibility, low-cost polymer synthesis [13][14], etc. Among several proposed RS mechanisms, conductive filament (CF) formation is well accepted in the scientific community with theoretical and experimental support [15][16][17]. There are mainly two types of CFs responsible for the RS in OSC-based devices. One is a carbon-rich filament due to the pyrolysis of the polymer [18][19][20], another one is due to the metal electrode atom migration through the polymer, which changes the resistive state of the device [21][22]. The metal filament formation is more common than the other and scientists focus more on this phenomenon because of low power consumption, long retention time, and good thermal stability [23][24].

Various OSCs have been investigated to elucidate the proper mechanism of CF formation and the switching mechanism. For example, Wang et al. demonstrate the Poly(3,4-ethylene-dioxythiophene): Poly (styrenesulfonate) (PEDOT: PSS) based RS using copper as top electrode (TE) [16], Cho et al. show the RS mechanism based on poly [(9,9-bis (6′-(N, N, N-trimethylammonium) hexyl)-2,7-fluorene)-co-(9,9-bis (2-(2- 2-methoxyethoxy) ethyl) fluorene) co (2,1,3benzothiadiazole)] dibromide due to conductive metal filament formation [17], Joo et al. show the cu ion drifting through secondary ion mass spectroscopy into the Poly(3-hexylthiophene-2,5-diyl) (P3HT), causing the RS in the device [25], Sun et al. explain the carbon nanotube-PEDOT: PSS based RS [26], Gao et al. demonstrate that the metal filament originates from the TE and reaches to the bottom electrode (BE), leading to RS in P3HT: [6,6]-phenyl C61-butyric acid methyl ester [27]. All these works focus mainly on an unpatterned bottom electrode; hence, how exactly the electric field and current are helping the RS to take place was never studied. The local field build-up at the highest points of the bottom electrode and lowest points of the top electrode can cause the RS, since the two electrodes are closer and hence the electric field strengths at these local points are much higher. To investigate the role of this factor, we have designed a perforated bottom electrode (PBE) instead of a traditional continuous metal electrode. The COMSOL Multiphysics simulation shows a high local electric field at the edges of the pattern, which is the reason to choose a patterned BE that consists of perforations. Two different shapes, the hexagonal and square patterns, are taken as the PBE to investigate the switching mechanism. It has been shown that the presence of Sulphur enhances the chance of metal filament formation, motivating us to choose the P3HT for this work [28],[29]. However, various types of RS mechanisms for the same material are proposed for the first time in this work under the presence of the PBE. There are mainly three different types of switching that have been observed in devices with PBE. First is the complete filament formation, which gives ohmic conduction at low resistive state (LRS), second is the mixture of incomplete filament formation, and space charge conduction (SCLC), and third is the inverted RS mechanism, where the device attains an intermediate OFF state after switching to LRS, followed by again attaining the same LRS state at lower voltage. The metal percolation is confirmed by the optical images and scanning electron microscope images, and the change in molecular arrangement with temperature as observed by the absorption spectroscopy. Different kinds of switching behavior in OSC devices will help to understand the underlying physics of charge carrier transport, metal ion movement, and molecular arrangement mechanisms that can be useful in upcoming artificial intelligence, neuromorphic computing, and memory devices.

## 2. Experimental section

### 2.1 Materials

P3HT, Chlorobenzene, and Aluminum (Al) beads were purchased from Sigma Aldrich. Si/SiO$_2$ substrates (100 nm) were purchased from University Wafer. All the materials are used without further purification.

### 2.2 Device fabrication

Si/SiO$_2$ substrates were sequentially cleaned under an ultrasonication process with acetone and isopropyl alcohol for 5 minutes in each solvent. An N$_2$ gun is used to dry the substrate. Finally, substrates were baked at 110 °C for 10 minutes. Lift-off resist (LOR) is spin-coated on the substrate at 6000 rpm for 45 seconds and baked at 90 °C for 2 minutes. After 2 minutes of cooling, a positive photoresist S1813 is spin-coated at 4500 rpm for 45 seconds. Lastly, 2 minutes of baking at 90 °C have been done. Lithography is performed using a photomask in a double-sided mask aligner (MJB6, SUSS MicroTec) under 15 seconds of exposure. The clear square and hexagonal patterns were achieved by 20 seconds of development in MF319. Then, Ti (03 nm) and Au (10 nm) were deposited by the DC magnetron sputtering method (AJA International, Inc/Orion Sputter PHASE) at 3 nm/minute and 9.8 nm/minute, respectively, in the presence of Argon gas at a pressure of 3 mtorr with the base pressure of $2 \times 10^{-7}$ torr. The substrates are then dipped into an NMP-based solution for 1.5 hours to lift off the unwanted metal. Finally, it is cleaned with IPA to get a proper PBE of Ti/Au on the substrate. P3HT of concentration 15 mg/ml was dissolved in chlorobenzene and stirred overnight for complete dissolution. The P3HT solution was spin-coated on the patterned substrate under the glove-box (M BRAUN, MB 20G) of N$_2$ atmosphere at 1000 rpm for 30 seconds and baked at 140 °C for 30 minutes to evaporate the solvent, resulting in the switching channel length of ~60 nm. Then, the TE of 100 nm Al is thermally evaporated using a metal mask at a rate of 0.5 Å/second at a pressure of $4 \times 10^{-6}$ mbar.

### 2.3 Characterization

The absorption spectrum is recorded by a UV-Vis-NIR Spectrometer - Lambda 950 (PerkinElmer). The lamella has been prepared for the cross-sectional transmission electron microscope (TEM) image by focused ion beam-scanning electron microscope (ThermoFischer Scientific, Helios 5 UC). The EDX spectrum of percolated metal into OSC is captured by the TEM (Thermo Scientific, Themis 300 G3). The current-voltage characteristics are measured with the Keysight Agilent B1500 semiconductor device analyzer.

## 3. Results and discussion

### 3.1 Device structure and simulation details:

Fig. 1(a) depicts the 3D schematic of the device. The substrate Si/SiO$_2$, perforated bottom Au electrode, switching P3HT channel, and Al TE are shown from the bottom to top of the figure. Fig. 1(b) shows the BE mask design with the hexagonal and square patterns. The hexagonal pattern of 120 μm arm length, which results in a 240 μm corner-to-corner distance, and a square pattern of 60 μm arm length, are shown in the zoomed-in pictures. 10 μm is the electrode width in both the structures. Fig. 1(c) shows the chemical structure of P3HT. COMSOL simulation is carried out to examine the electric field distribution between the perforated BE and the TE at 10 V, the result is shown in Fig. 1 (d). The semiconductor material module is used to perform the simulation. The drift-diffusion equations are used in the simulation, which were mentioned in our previous work [30], [31]. The basic properties of the OSC used in the simulation are listed in Table 1 in the supporting information. The device geometry and corresponding mesh structure are shown in Fig. S1. A high electric field is observed at the edges of the source electrode structure (red region), as shown in Fig. 1 (d) at 10 V. Non-conventional paths of charge carriers would be generated due to this non-uniform electric field. Moreover, this electric field enhances the probability of filament formation, which is the main reason to choose the perforated source electrode in this work.

### 3.2 Complete filament formation:

The current-voltage (I-V) characteristics of the device with a square PBE pattern are shown in Fig. 2(a). The single cycle with sweeping direction from 0 V to +20 V, then back to -20 V through 0 V, and lastly from -20 V to 0 V is shown in Fig. 2(b). The LRS is achieved in the 2$^{nd}$ step at a voltage of 3.7 V. This phenomenon is known as the 'SET' process. It remains in the LRS in the negative bias region. LRS turns into high resistive state (HRS) at a bias of -6.9 V in the 3$^{rd}$ step of the characteristics, is known as the 'RESET' process. Therefore, it shows bipolar RS behavior. The log-log plot of the positive bias regime is shown in Fig. 2(c). Primarily, current increases with applied voltage with a slope of 0.88, indicating the ohmic conduction. The slope increases to 2.60 with increasing the bias indicates the space charge limited current (SCLC) regime.

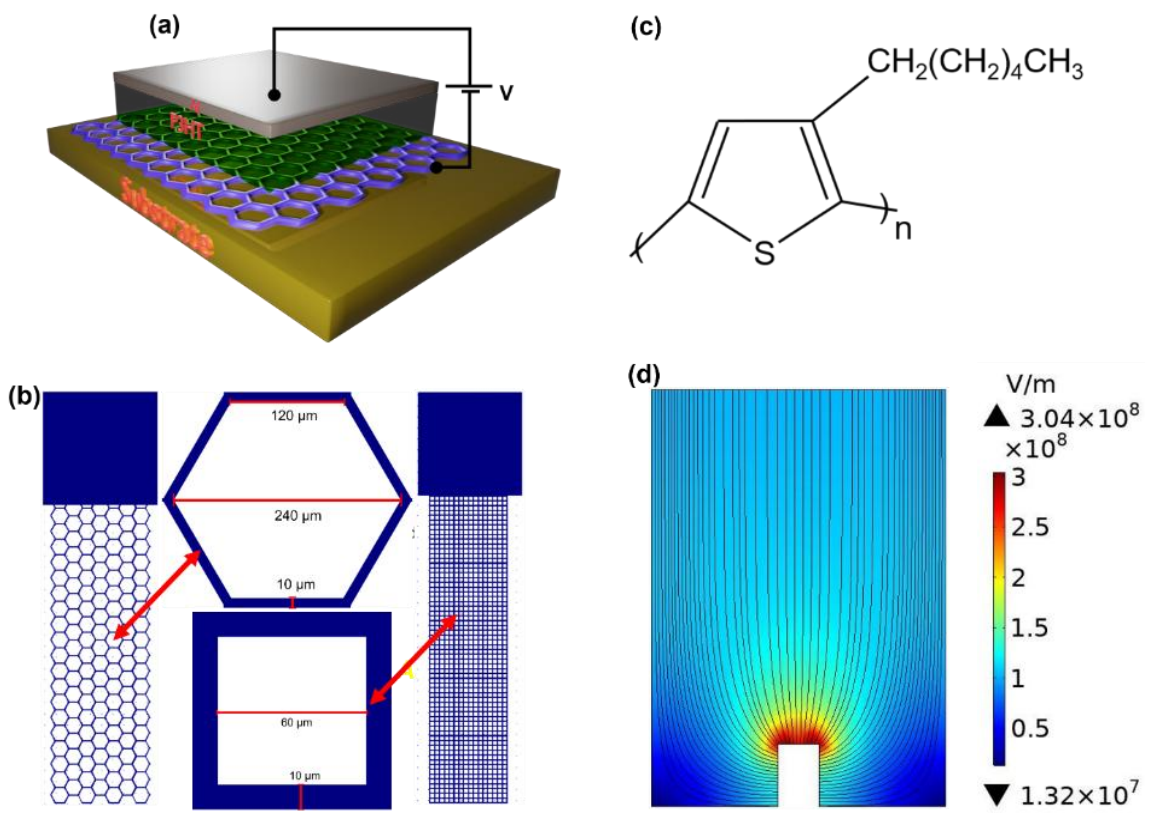

**Figure 1.** (a) 3D schematic of the device. (b) PBE structures of a hexagonal and square pattern. (c) Chemical structure of P3HT. (d) Electric field profile at 10 V using COMSOL simulation.

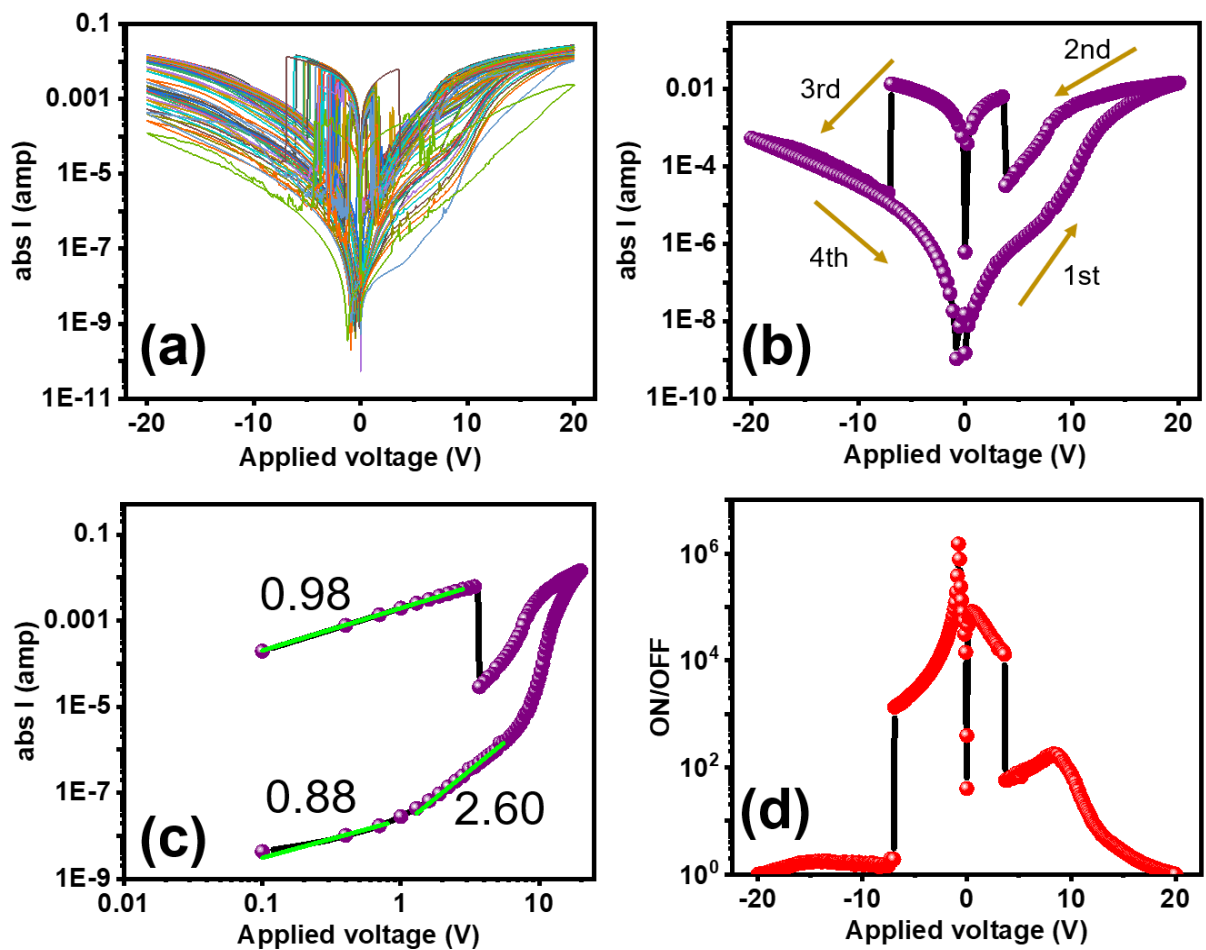

**Figure 2.** (a) I-V characteristics of the device for multiple cycles. (b) Single cycle of the device. Starts from 0 V to 20 V and then sweeping back to -20 V through 0 V and finally sweeps back to 0 V. (c) Log-log scale of the positive cycle (0 V to 20 V to 0 V) with fitted slope of (b). (d) ON/OFF ratio with respect to applied voltage calculated from (b).

The plot is fitted after the "SET" process yields the slope of 0.98, might be the metallic filament formation, which is reached at the BE. The Highest ON/OFF ratio of $10^6$ is achieved in this device at a bias of -0.8 V. The long memory window of -7 V to 3.6 V helps to distinguish each resistance of the device, as shown in Fig. 2(d). The metallic filament formation is the primary reason behind this switching, which can be confirmed by the optical and transmission electron microscope (TEM) images, as discussed in Fig. 3. Fig. 3(a) shows an optical image from the top view of the device before applying any bias. It shows only Al from the top side, where the P3HT and perforated Au pattern are beneath the Al. On the other hand, Fig. 3(b) shows the optical image of the same region after measuring the current-voltage characteristics. The metal percolation due to the electric field is clearly visible in this image. The device is cut to get a cross-section along one arm of the square by a focused ion beam after multiple cycles. The marking area for the lamella preparation of a square arm (10 μm width) is shown in Fig. S3. Hence, the EDX spectrum (cross-sectional TEM) at LRS is analyzed, as shown in Fig. 3(c), (d). The scanning region from Al to Si is shown in Fig. 3(c), and the corresponding atomic fraction of Al and Au (other elements are not shown here) is shown in Fig. 3(d). The overlapping region of Al and Au clearly signifies the Al filament formation in the device, which touches the bottom Au electrode. This filament formation is responsible for the observed RS in the device. Hence, the rest of the discussion in this paper will be based on metal filament formation. The proposed mechanism of this switching is illustrated in Fig. 4. Fig. 4(a) shows the virgin two-terminal device with no percolation of metal. The anode material (TE) is oxidized to an ion under the applied positive bias. These ions move through the channel and reach to the BE. Now, at this electrode, all these ions are transferred into atoms and make a conductive filament from the cathode (BE) to the anode (TE) according to electrochemical metallization theory [32]. However, in this case, the anode material (Al) is oxidised to $Al^{n+}$ (n = 1, 3) and deionised to an atom by receiving the electrons from the BE as shown in Fig. 4(b). This contradicts the proposed electrochemical metallization theory. This theory is well applicable in the traditional solid-state electrolytes such as AuS, CuS, etc., those have high diffusion coefficients of metal ions. As the organic layer has lower ionic mobility, the Al ions are easily turned into atoms after covering a short distance in the channel [21][27]. After the multiple oxidation/reduction processes, the Al atoms precipitate from the TE and extend to the BE as shown in Fig. 4(c), resulting in LRS. At the negative voltage, weakest part of the filament is broken up due to joule heating, causing to increase in the resistance of the channel, hence, HRS is again set, as shown in Fig. 4(d) [27].

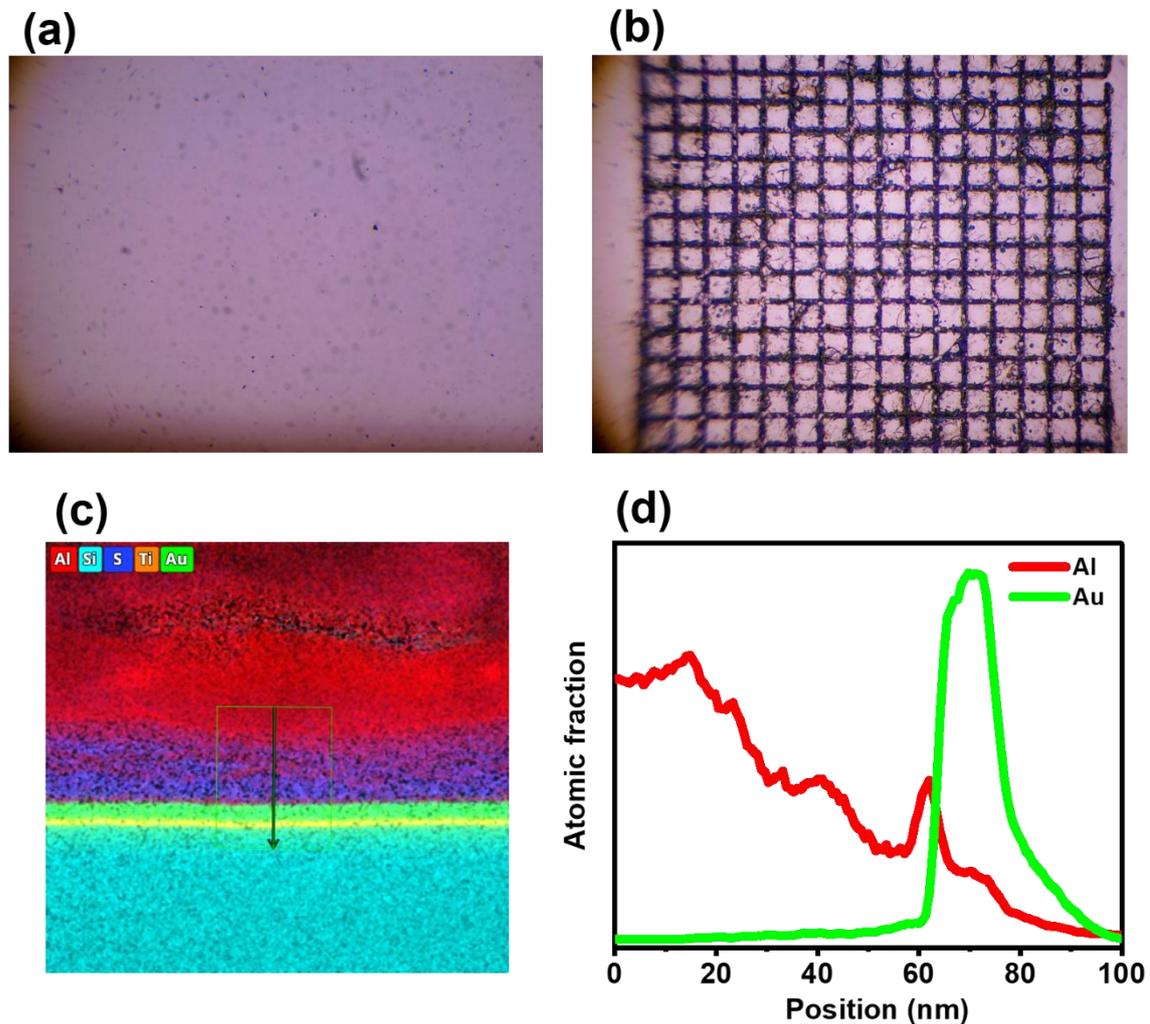

**Figure 3.** Optical image (top view) (a) fresh device (before applying any bias), (b) after multiple cycles. (c) and (d) EDX spectrum of the device after multiple cycles shows the top metal pecolation through OSC and touches the Au bottom layer.

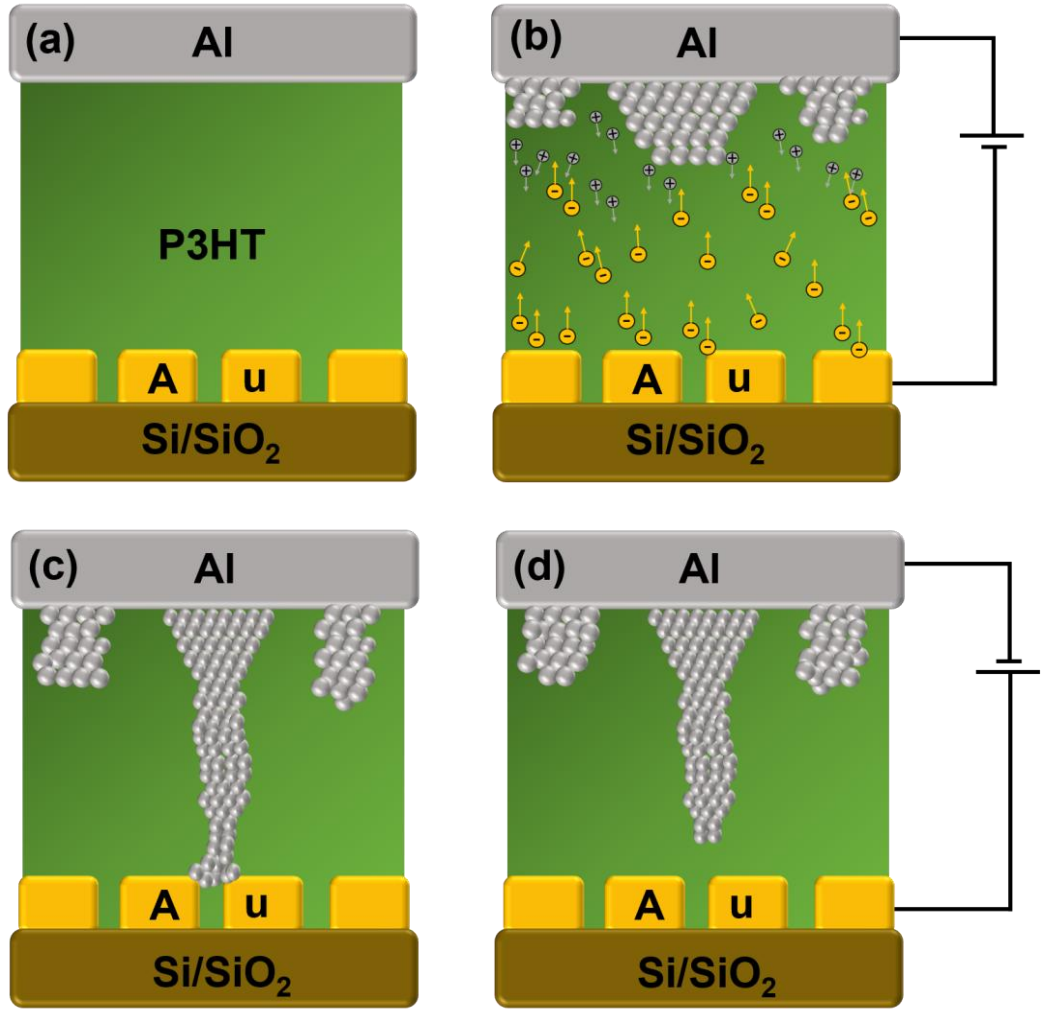

**Figure 4.** (a) Virgin device with distinct interfaces. (b) Oxidation and reduction process of Al atoms under the positive bias. (c) Stable Al filament reaches the PBE. (d) Filament breaks at negative bias result in back to HRS.

**3.3 Unstable filament formation:**

In this case, the hexagonal perforated pattern is considered as BE. I-V characteristics of the device are shown in Fig. 5(a). SET process is achieved at 10.4 V while back to initial HRS is at -16.8 V, as shown in Fig. 5(b). The cumulative probability for 50 cycles for the SET and RESET voltage is shown in Fig. 5(c). The average values for the SET and RESET processes are 12.9 V and -14.0 V, respectively. The SET voltage shows a wider (2.4 V to 19.2 V) distribution than the RESET voltage (-18.2 V to -10.2 V), indicating the metal filament formation responsible for the resistive switching. There is a competition among different filamentary paths to form, resulting in a larger distribution of SET voltage. Conspicuously, rupture of the existing filament is not random like the formation process, leading to a narrower threshold voltage range distribution. Many previous reports validated these phenomena in the case of a switching device [27][33][34]. Therefore, this might be a clue for the support of filament formation in our device. The log-log plot is shown in Fig. 5(d). Ohmic conduction is observed at the low bias region with a slope of 0.81. SCLC region with a slope of 2.82 is maintained at the high bias region. After achieving the LRS, the slope at high bias is 1.34, signifying the filament instability in the device. By decreasing the bias, the slope increases toward 2.0, indicating that the filament suppression and SCLC dominating current in the device. A long voltage window (-16.8 V to 11.6 V) provides the memory region having ON/OFF ratio ~$10^3$, as shown in Fig. 5(e). Fig. 6(a) shows the optical image of the device after multiple cycles. This image depicts the metal percolation in the channel. However, Fig. 3(b) shows more compact, continuous, and complete percolation in the square pattern compared to this hexagonal structure, supporting the instability of filament formation. The more electrical field interaction in the case of square pattern due to lower angle between the neighbouring edges (90°) compared to the hexagonal (120°), might be the reason for complete metal percolation in the square structure. Fig. 6(b) and (c) show the virgin device and filament formation by oxidation/reduction process of Al atoms under the positive bias at the TE. In this case, the filament is not fully grown to touch the BE; hence, no complete metallic behavior is observed after the SET process in the log-log plot. However, the effective distance becomes less, as shown in Fig. 6(d), which causes to reach the LRS (SET). The SCLC dominates in this case even after achieving the LRS, as shown in Fig. 6 (e). Most of the work reports the complete metallic behavior after SET. However, a new kind of phenomenon is observed here, which consists of both a complete metallic formation like in the previous case and a dominance of SCLC transport mechanism over the metallic conduction. The filament breaks at negative bias leading back to HRS, as shown in Fig. 6(f).

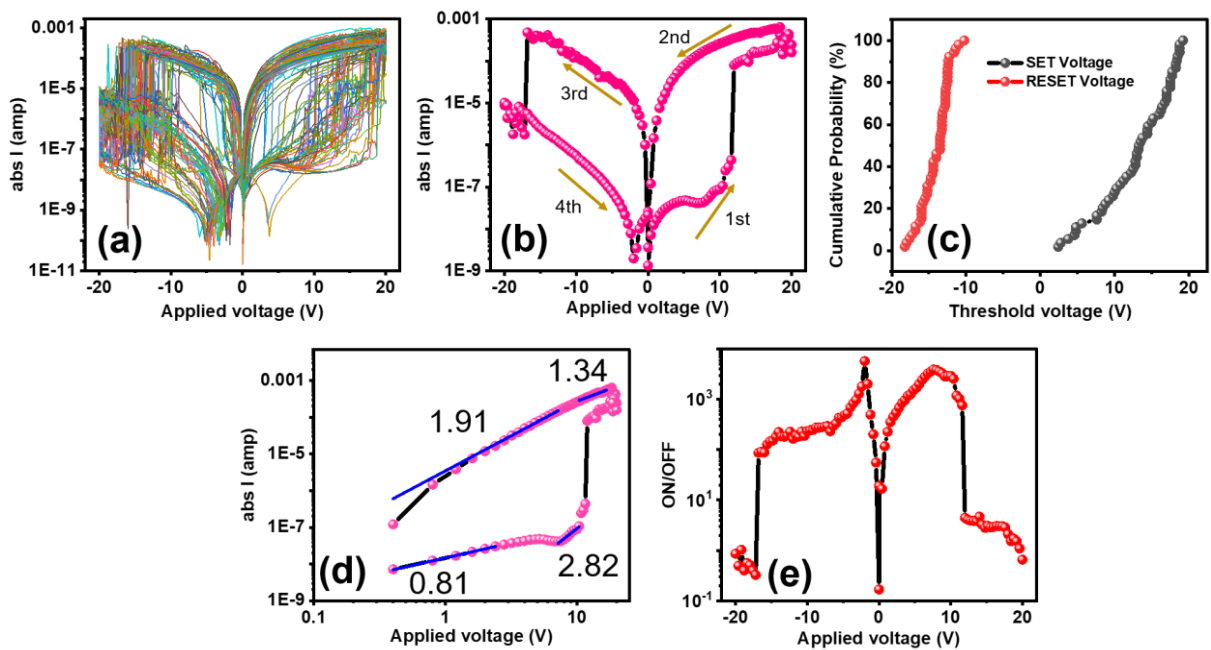

**Figure 5.** (a) I-V characteristics of the device for multiple cycles. (b) Single cycle of the device. Starts from 0 V to 20 V and then sweeping back to -20 V through 0 V and finally sweeps back to 0 V. (c) Cumulative probability of threshold voltage. (d) Log-log scale of the positive cycle (0 V to 20 V to 0 V) with fitted slope of (b). (e) ON/OFF ratio with respect to applied voltage calculated from (b).

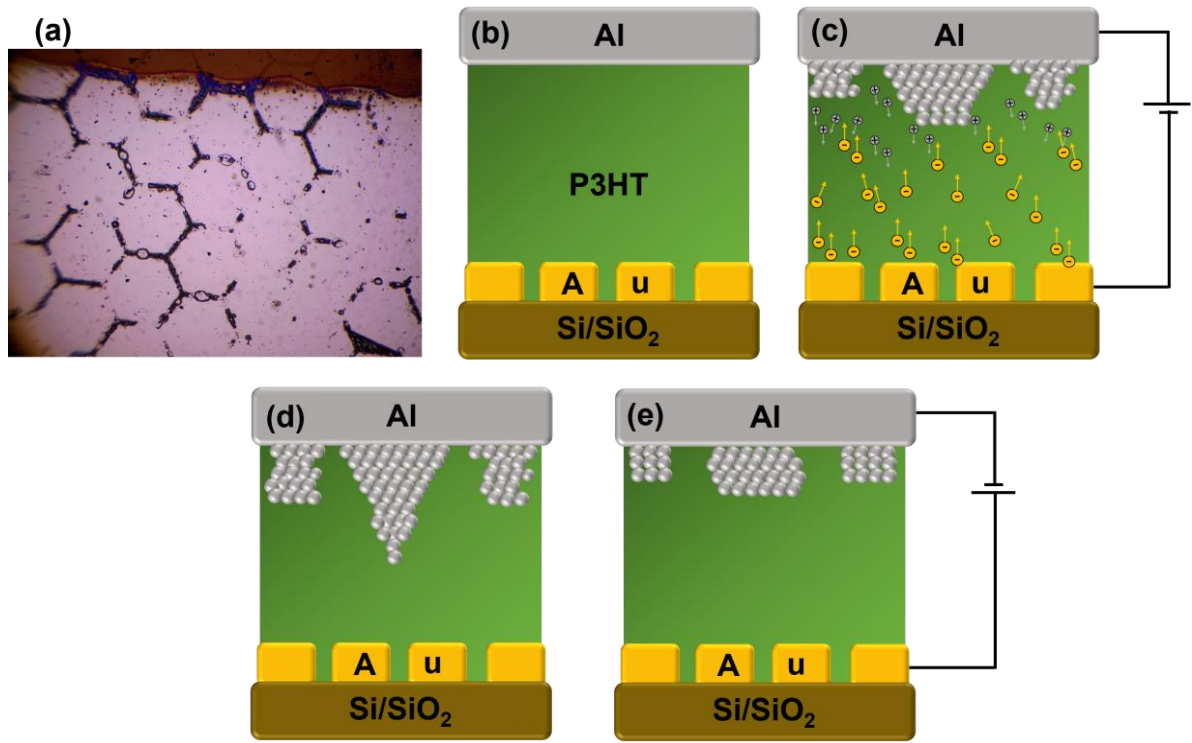

**Figure 6.** (a) Optical image (top view) of the device consists of hexagonal PBE after multiple cycles. (b) Virgin device with distinct interfaces. (c) Oxidation and reduction process of Al atoms under the positive voltage. (d) Formation of a partial filament. (e) Filament breaks at negative voltage result in back to HRS.

**3.4 Resistive switching with morphological changes of OSC:**

In this section, we will discuss another type of switching in the device. Fig. 7(a) shows the RS for 50 cycles of the device from 0 V to +20 V to 0 V to -20 V to 0 V. Fig. 7(b) shows a single cycle that consists of a sharp switching (HRS to LRS) followed by an inverted switching (LRS to an intermediate OFF state). The SCLC with low voltage and instability of the filament after the "SET" process are also observed in this mechanism, as shown in Fig. 7(c). The detailed mechanism of this switching is illustrated in Fig. 8. Firstly, the sharp switching from HRS to LRS under positive voltage (1$^{st}$ step in Fig. 7(b)) is due to partial filament formation by the oxidation/reduction of TE atoms and SCLC phenomenon, as discussed earlier, shown in Fig. 8(a-d). The molecular structure of P3HT changes from crystalline to amorphous due to local joule heating at high voltage, as shown in Fig. 8(e). The hexagonal-shaped perforation creates a high local electric field region, causing high local current paths. The high current in the device leads to a rise in temperature of the local regions of the OSC are responsible for the morphological changes that follows by the charge carriers, results to an intermediate OFF state (2$^{nd}$ step in Fig. 7(b)), i.e inverted switching, as discussed by Ramesh et al. [35]. The tendency of morphological changes of P3HT can be seen from the absorption spectroscopy, as shown in Fig. 7(d). In case of less ordered OSCs, a lower energy gap between the highest occupied molecular orbital and lowest unoccupied molecular orbital has been observed due to minimum intermolecular interaction compared to crystalline polymer. Hence, a blue shift is expected in the absorption spectroscopy for randomly oriented or less ordered state of a molecule [36]. The broadening of the absorption peak along with the blue shift signifies the decrease of the crystallinity of P3HT with increasing temperature. Therefore, this absorption spectrum supports the change in molecular arrangement with local heating as observed in the switching characteristics. However, the charge carriers follow the fresh crystalline paths while sweeping back from the high voltage, which leads to LRS (4$^{th}$ step in Fig.7(b)) due to less current, as shown in Fig. 8(f). At negative bias region, no sharp transition from LRS to HRS has been observed. The slow dissolution of the atoms could be the reason for this gradual switching, as shown in Fig. 8(g). However, there are a few cycles that show a sharp change to HRS, as shown in Fig. S4(a). Fig. S4(b) shows the single cycle in which the transition from LRS to HRS is at -17.2 V. The sudden filament break like section 3.3 might be the reason for this, as shown in Fig. S5(g).

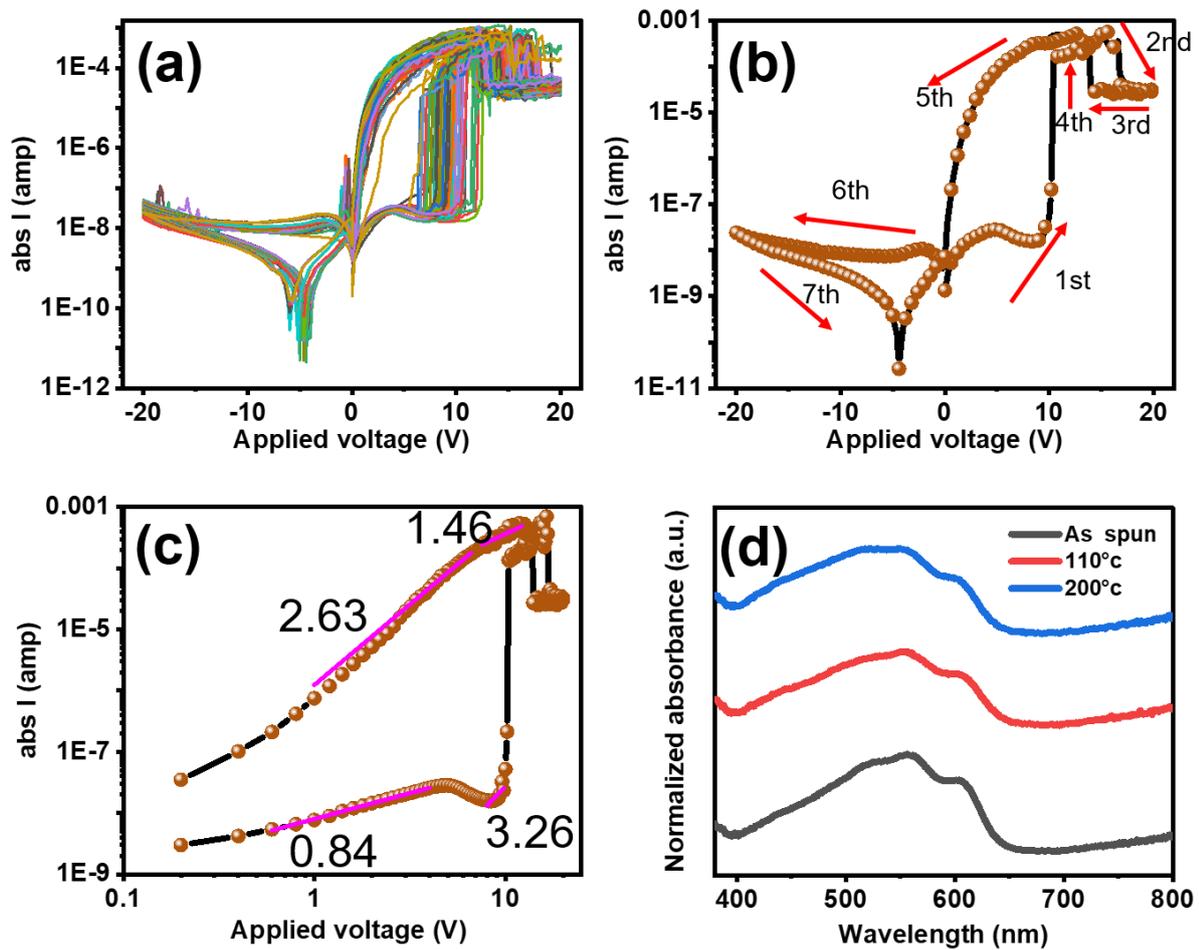

**Figure 7.** (a) I-V characteristics of the device for multiple cycles. (b) Single cycle of the device. Starts from 0 V to 20 V and then sweeping back to -20 V through 0 V and finally sweeps back to 0 V. (c) Log-log scale of the positive cycle (0 V to 20 V to 0 V) with fitted slope of (b). (d) Absorbance spectroscopy of spin coated OSC at different temperature.

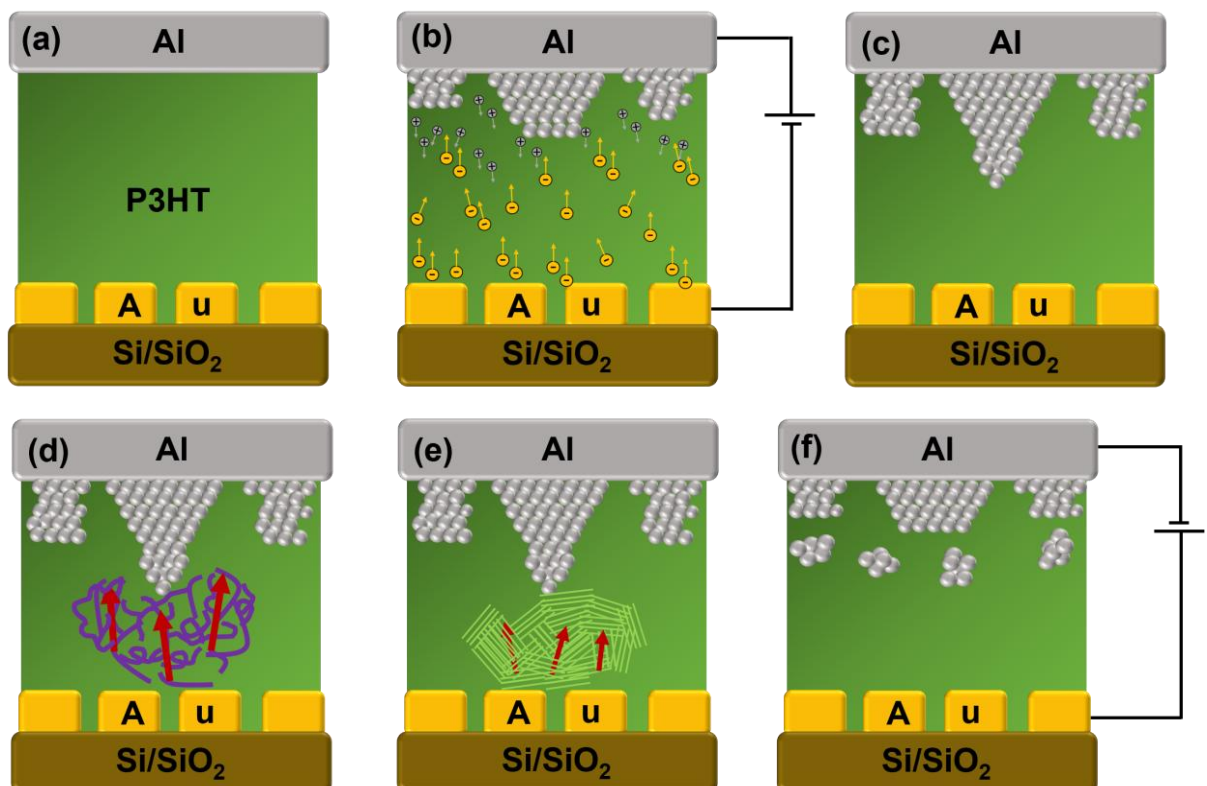

**Figure 8.** (a) Virgin device with distinct interfaces. (b) Oxidation and reduction process of Al atoms under the positive voltage. (c) Formation of partial filament. (d) Crystalline to amorphous due to joule heating leads to inverted switching. (e) Carriers follow the crystalline path again while sweeping back leads to LRS. (f) Slow dissolution of the filament causes gradual switching.

## 4. Conclusions

In conclusion, we successfully designed and fabricated the RS devices, viz., perforated Au/P3HT/Al. Different types of switching mechanisms have been discussed in this work. The complete, unstable filament formation, along with the change in molecular arrangement, are the main reasons behind these characteristics. In the first case, the complete filament is formed between the electrodes, shows metal-like behavior after the SET process. In the second case, a complete stable filament is absent, where charge conduction takes place via SCLC. Hence, switching occurs due to a decrease in the effective length from TE to BE by forming an incomplete filament. Lastly, the inverted switching is also observed after reaching LRS from HRS. The change in molecular arrangement of OSC due to joule heating at high current in the device causes the sudden drop of current from LRS to an intermediate OFF state current. Understanding the origin of different kinds of switching helps to put applications in the field of memory and nano-electronics.

**Data availability**

The data that supports the findings of this study are available on request from the authors.

**Conflict of interest**

The authors declare no known conflicts of interest related to this work.

**Acknowledgments**

SG thanks the Indian Institute of Technology (IIT) Bombay for the fellowship. The authors thank the Centre of Excellence in Nanoelectronics (CEN), and the National Centre for Photovoltaic Research and Education (NCPRE) of IIT Bombay for using the facilities of fabrication and device characterization. The authors will also thank the Sophisticated Analytical Instrument Facility (SAIF) and the Centre for Sophisticated Instruments and Facilities (CSIF) of IIT Bombay for the lamella preparation and TEM imaging.

**References**

[1] B. W. Zhang *et al.*, "Lead-Free Perovskites and Metal Halides for Resistive Switching Memory and Artificial Synapse," *John Wiley and Sons Inc*, vol. 5, no. 6, Jun. 01, 2024, doi: 10.1002/sstr.202300524.

[2] X. Liu *et al.*, "Flexible Transparent High-Efficiency Photoelectric Perovskite Resistive Switching Memory," *Adv Funct Mater*, vol. 32, no. 38, Sep. 2022, doi: 10.1002/adfm.202202951.

[3] K. Yang *et al.*, "Anatomy of resistive switching behavior in titanium oxide based RRAM device," *Mater Sci Semicond Process*, vol. 143, Jun. 2022, doi: 10.1016/j.mssp.2022.106492.

[4] P. H. Chen, C. Y. Hsieh, and H. Y. Yang, "Effects of Charge Quantity Induced by Different Forming Methods in Solid Electrolyte GeSO-Based Resistance Switching Device with Copper Electrode," *IEEE Trans Electron Devices*, vol. 67, no. 6, pp. 2324–2328, Jun. 2020, doi: 10.1109/TED.2020.2985084.

[5] S. J. Choi *et al.*, "In situ observation of voltage-induced multilevel resistive switching in solid electrolyte memory," *Advanced Materials*, vol. 23, no. 29, pp. 3272–3277, Aug. 2011, doi: 10.1002/adma.201100507.

[6] J. J. Ryu, K. Jeon, T. Eom, M. K. Yang, H. Sohn, and G. H. Kim, "Optimized chalcogenide medium for inherently activated resistive switching device," *Appl Surf Sci*, vol. 641, Dec. 2023, doi: 10.1016/j.apsusc.2023.158444.

[7] H. Yu, P. K. Zhou, and X. Chen, "Intramolecular Hydrogen Bonding Interactions Induced Enhancement in Resistive Switching Memory Performance for Covalent Organic Framework-Based Memristors," *Adv Funct Mater*, vol. 33, no. 44, Oct. 2023, doi: 10.1002/adfm.202308336.

[8] P. K. Zhou, H. Yu, Y. Li, H. Yu, Q. Chen, and X. Chen, "Recent advances in covalent organic polymers-based thin films as memory devices," Apr. 15, 2024, *John Wiley and Sons Inc*. doi: 10.1002/pol.20230273.

[9] Z. J. Donhauser, B. A. Mantooth, K. F. Kelly, L. A. Bumm, J. D. Monnell, J. J. Stapleton, D. W. Price Jr., A. M. Rawlett, D. L. Allara, J. M. Tour, and P. S. Weiss, "Conductance Switching in Single Molecules Through Conformational Changes," *Science*, vol. 292, no. 5525, pp. 2303-2307, June. 2001, doi: 10.1126/science.1060294.

[10] D. I. Son *et al.*, "Bistable organic memory device with gold nanoparticles embedded in a conducting poly (N -vinylcarbazole) colloids hybrid," *Journal of Physical Chemistry C*, vol. 115, no. 5, pp. 2341–2348, Feb. 2011, doi: 10.1021/jp110030x.

[11] L. D. Bozano, B. W. Kean, V. R. Deline, J. R. Salem, and J. C. Scott, "Mechanism for bistability in organic memory elements," *Appl Phys Lett*, vol. 84, no. 4, pp. 607–609, Jan. 2004, doi: 10.1063/1.1643547.

[12] J. J. Yang, M. D. Pickett, X. Li, D. A. A. Ohlberg, D. R. Stewart, and R. S. Williams, "Memristive switching mechanism for metal/oxide/metal nanodevices," *Nat Nanotechnol*, vol. 3, no. 7, pp. 429–433, 2008, doi: 10.1038/nnano.2008.160.

[13] S. Ghosh, G. Shukla, and P. Kumar, "Artificial short term synaptic behavior of organic polymer device capable of detecting both visible and infrared signals," *Mater Today Commun*, vol. 47, Jul. 2025, doi: 10.1016/j.mtcomm.2025.113115.

[14] H. Bronstein, C. B. Nielsen, B. C. Schroeder, and I. McCulloch, "The role of chemical design in the performance of organic semiconductors," *Nature Research*, vol. 4, no. 2, pp. 66-77, Feb. 01, 2020, doi: 10.1038/s41570-019-0152-9.

[15] D. Prime and S. Paul, "Overview of organic memory devices," *Royal Society*, Oct. 28, 2009, Vol. 367, no. 1905, pp. 4141-4157, doi: 10.1098/rsta.2009.0165.

[16] Z. Wang, F. Zeng, J. Yang, C. Chen, and F. Pan, "Resistive switching induced by metallic filaments formation through poly(3,4-ethylene-dioxythiophene): Poly(styrenesulfonate)," *ACS Appl Mater Interfaces*, vol. 4, no. 1, pp. 447–453, Jan. 2012, doi: 10.1021/am201518v.

[17] B. Cho, J. M. Yun, S. Song, Y. Ji, D. Y. Kim, and T. Lee, "Direct observation of Ag filamentary paths in organic resistive memory devices," *Adv Funct Mater*, vol. 21, no. 20, pp. 3976–3981, Oct. 2011, doi: 10.1002/adfm.201101210.

[18] BH Lee, H Bae, H Seong, DI Lee, H Park, YJ Choi, SG Im, S.G., SO Kim, and YK Choi, 2015. "Direct observation of a carbon filament in water-resistant organic memory," *ACS nano*, vol. 9, no. 7, pp.7306-7313, Jul. 2015, doi: 10.1021/acsnano.5b02199.

[19] N. Raeis-Hosseini and J. S. Lee, "Controlling the Resistive Switching Behavior in Starch-Based Flexible Biomemristors," *ACS Appl Mater Interfaces*, vol. 8, no. 11, pp. 7326–7332, Mar. 2016, doi: 10.1021/acsami.6b01559.

[20] Y. Sun, D. Wen, and F. Sun, "Multistage Resistive Switching Processes in Filament Conduction-Based Organic Memory Devices," *J Electrochem Soc*, vol. 167, no. 2, p. 027529, Jan. 2020, doi: 10.1149/1945-7111/ab6b0a.

[21] Q. Liu *et al.*, "Real-time observation on dynamic growth/dissolution of conductive filaments in oxide-electrolyte-based ReRAM," *Advanced Materials*, vol. 24, no. 14, pp. 1844–1849, Apr. 2012, doi: 10.1002/adma.201104104.

[22] S. H. Lee, H. L. Park, M. H. Kim, M. H. Kim, B. G. Park, and S. D. Lee, "Realization of Biomimetic Synaptic Functions in a One-Cell Organic Resistive Switching Device

Using the Diffusive Parameter of Conductive Filaments," *ACS Appl Mater Interfaces*, vol. 12, no. 46, pp. 51719–51728, Nov. 2020, doi: 10.1021/acsami.0c15519.

[23] S. Mö Ller, C. Perlov, W. Jackson, C. Taussig, and S. R. Forrest, "A polymer/semiconductor write-once read-many-times memory," *Nature*, vol. 426, no. 6963, pp. 166-169, 2003, doi: 10.1038/nature02070.

[24] S. Wu *et al.*, "A polymer-electrolyte-based atomic switch," *Adv Funct Mater*, vol. 21, no. 1, Jan. 2011, pp. 93–99, doi: 10.1002/adfm.201001520.

[25] WJ Joo, TL Choi, KH Lee, and Y Chung, 2007. "Study on threshold behavior of operation voltage in metal filament-based polymer memory," *The Journal of Physical Chemistry B*, vol. 111, no. 27, pp.7756-7760, Jul. 2007, doi: 10.1021/jp0649899.

[26] Y. Sun, L. Li, D. Wen, X. Bai, and G. Li, "Bistable electrical switching and nonvolatile memory effect in carbon nanotube-poly(3,4-ethylenedioxythiophene): poly(styrenesulfonate) composite films," *Physical Chemistry Chemical Physics*, vol. 17, no. 26, pp. 17150–17158, Jul. 2015, doi: 10.1039/c5cp02164b.

[27] S Gao, C Song, C Chen, F Zeng, and F Pan, 2012. "Dynamic processes of resistive switching in metallic filament-based organic memory devices," *The Journal of Physical Chemistry C*, vol. 116, no. 33, pp.17955-17959, Aug. 2012, doi: 10.1021/jp305482c.

[28] W. J. Joo *et al.*, "Metal filament growth in electrically conductive polymers for nonvolatile memory application," *Journal of Physical Chemistry B*, vol. 110, no. 47, pp. 23812–23816, Nov. 2006, doi: 10.1021/jp0649899.

[29] N. Knorr, R. Wirtz, S. Rosselli, and G. Nelles, "Field-absorbed water induced electrochemical processes in organic thin film junctions," *Journal of Physical Chemistry C*, vol. 114, no. 37, pp. 15791–15796, Sep. 2010, doi: 10.1021/jp103625w.

[30] S. Ghosh, R. Singh Bisht, and P. Kumar, "Simulation Study on Comparison of 'Inside-Channel' and 'On-Dielectric' Source Contact Modifications on the Performance of the Vertical Organic Field Effect Transistors," *IEEE Trans Electron Devices*, vol. 71, no. 10, pp. 6293–6298, 2024, doi: 10.1109/TED.2024.3442188.

[31] S. Ghosh, A. D. Laishram, and P. Kumar, "Simulation on the Miniaturization and Performance Improvement Study of Gr/MoS2 Based Vertical Field Effect Transistor," *Adv Theory Simul*, Sep. 2025, doi: 10.1002/adts.202500139.s

[32] R. Waser, R. Dittmann, C. Staikov, and K. Szot, "Redox-based resistive switching memories nanoionic mechanisms, prospects, and challenges," Jul. 13, 2009. doi: 10.1002/adma.200900375.

[33] Y. C. Yang, F. Pan, Q. Liu, M. Liu, and F. Zeng, "Fully room-temperature-fabricated nonvolatile resistive memory for ultrafast and high-density memory application," *Nano Lett*, vol. 9, no. 4, pp. 1636–1643, Apr. 2009, doi: 10.1021/nl900006g.

[34] Z. Wang, P. B. Griffin, J. McVittie, S. Wong, P. C. McIntyre, and Y. Nishi, "Resistive switching mechanism in ZNxCd1-xS nonvolatile memory devices," *IEEE Electron Device Letters*, vol. 28, no. 1, pp. 14–16, Jan. 2007, doi: 10.1109/LED.2006.887640.

[35] R. S. Bisht and P. Kumar, "Inverted resistive switching mechanism in polycrystalline PBTTT-C14 polymer devices based on contact geometry and molecular packing for neuromorphic memory," *J Mater Chem C Mater*, vol. 13, no. 26, pp. 13404–13414, May 2025, doi: 10.1039/d5tc01372k.

[36] H. Proehl, T. Dienel, R. Nitsche, and T. Fritz, "Formation of solid-state excitons in ultrathin crystalline films of PTCDA: From single molecules to molecular stacks," *Phys Rev Lett*, vol. 93, no. 9, Aug. 2004, doi: 10.1103/PhysRevLett.93.097403.

# Supporting Information

## Resistive-Switching Dynamics in Poly(3-hexylthiophene-2,5-diyl) Thin Films under Perforated Bottom Electrode

*Sirsendu Ghosh and Pramod Kumar\**

Department of Physics, Indian Institute of Technology Bombay, Mumbai, Maharashtra, 400076, India

*E-mail: pramod_k@iitb.ac.in

Table 1. Basic properties of P3HT [1]:

| Relative permittivity, $\varepsilon_r$ | Band gap, $E_{g0}$ [eV] | Electron affinity, $\chi_0$ [eV] | Effective density of states, valance band, $N_v$ [cm$^{-3}$] | Effective density of states, conduction band, $N_c$ [cm$^{-3}$] | Electron mobility, $\mu_n$ [cm$^2$ V$^{-1}$ s$^{-1}$] | Hole mobility, $\mu_p$ [cm$^2$ V$^{-1}$ s$^{-1}$] |
|---|---|---|---|---|---|---|
| 3 | 2 | 3.20 | $1 \times 10^{20}$ | $1 \times 10^{20}$ | $1 \times 10^{-4}$ | $1 \times 10^{-4}$ |

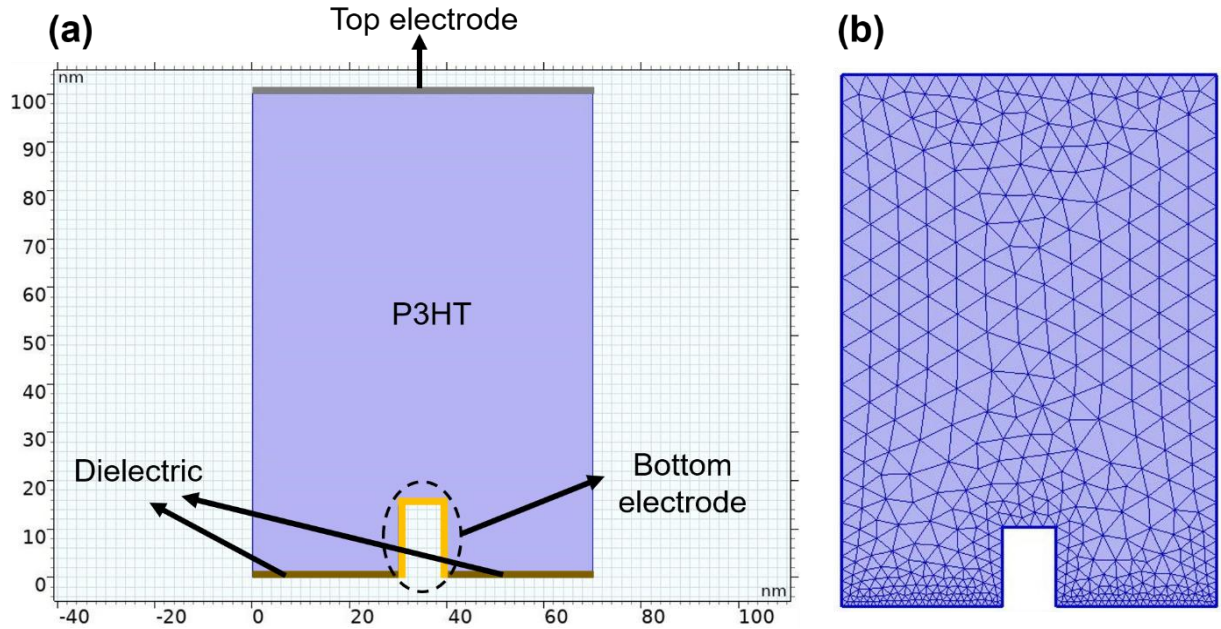

**Figure S1.** (a) Simulation structure of the device. Bottom electrode height and width are 15 and 10 nm, respectively. Perforated length 60 nm across the bottom electrode and the channel is 100 nm. Dielectric thickness is 100 nm. (b) Mesh of the device structure to perform the simulation.

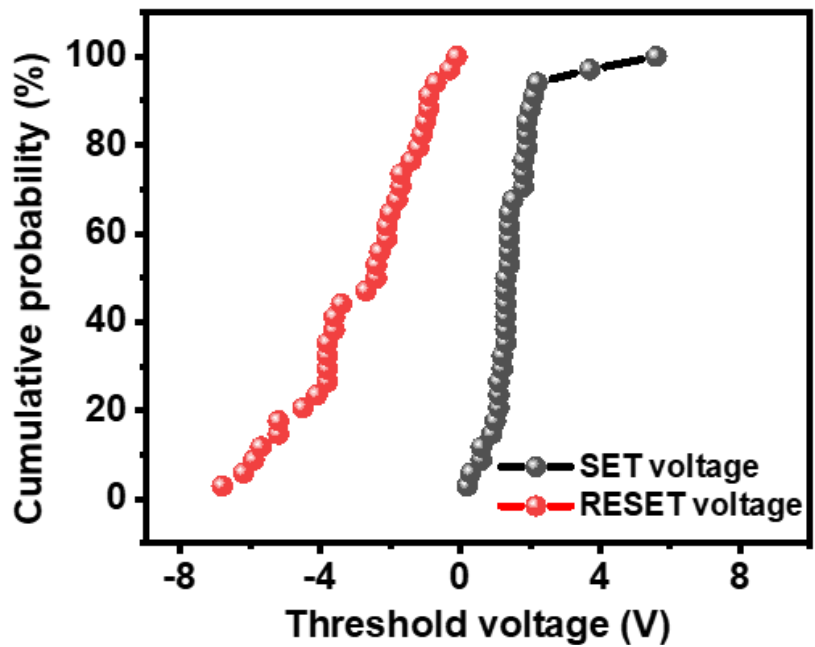

**Figure S2.** Cumulative probability of threshold voltage for the device having complete metal filament formation with square perforated bottom electrode structure. The average threshold value for the 'SET' and 'RESET' process is 1.5 V and -2.8 V, respectively.

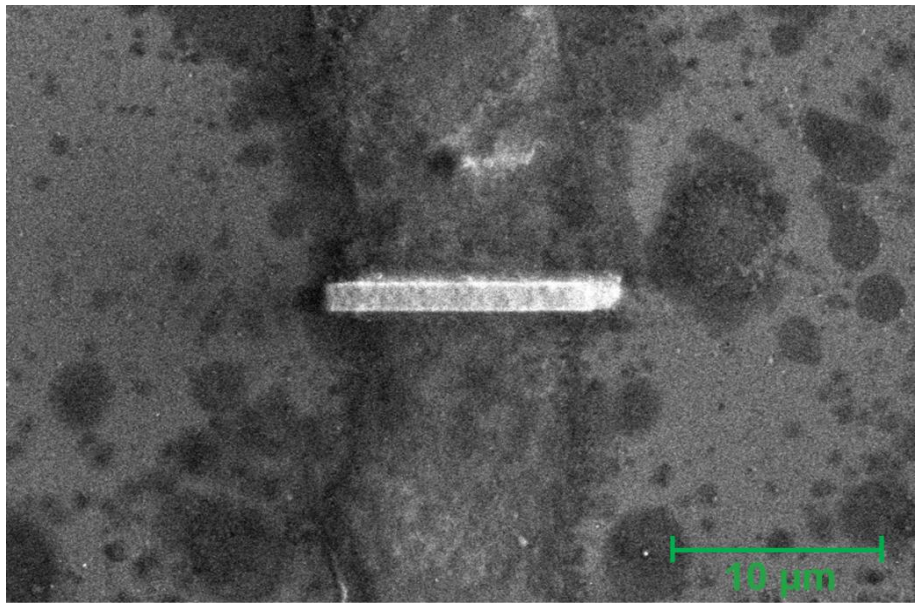

**Figure S3.** The marking area is chosen for the lamella preparation to perform the cross-sectional TEM. The width of the arm (cutting part) is 10 μm.

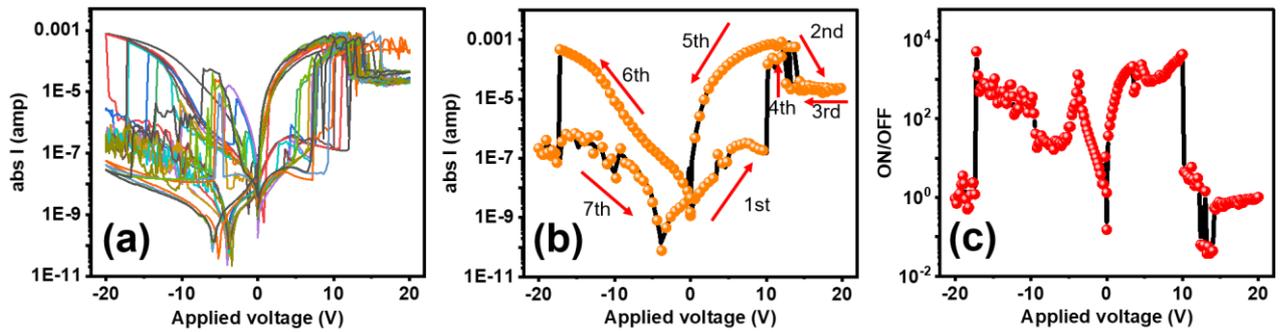

**Figure S4.** (a) I-V characteristics of the device for 10 cycles. (b) Single cycle of the device. Starts from 0 V to 20 V and then sweeping back to -20 V through 0 V and finally sweeps back to 0 V. (c) Variation of ON/OFF ratio with applied voltage calculated from (b).

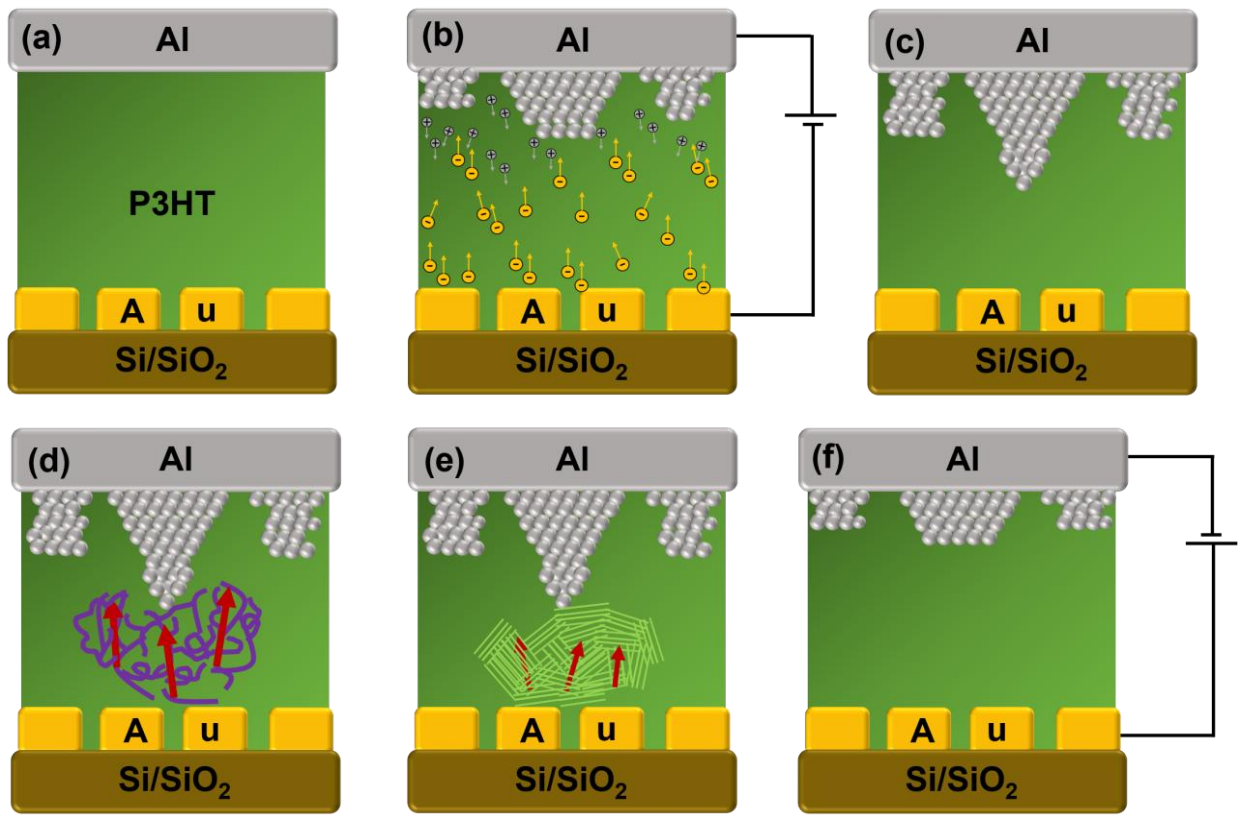

**Figure S5.** (a) Virgin device with distinct interfaces. (b) Oxidation and reduction process of Al atoms under the positive voltage. (c) Formation of partial filament. (d) Crystalline to amorphous due to joule heating leads to inverted switching. (e) Carriers follow the crystalline path again while sweeping back leads to LRS. (f) Sudden filament break causes sharp switching at negative bias.

**Reference:**

[1]   I. Alam, R. Mollick, and M. A. Ashraf, "Numerical simulation of Cs2AgBiBr6-based perovskite solar cell with ZnO nanorod and P3HT as the charge transport layers," *Physica B Condens Matter*, vol. 618, Oct. 2021, doi: 10.1016/j.physb.2021.413187.